\documentstyle[11pt,paspconf,epsf]{article}

\begin{document}

\title{Observations of Candidate $z\sim1.54$ Quasar Host Clusters}
\author{Patrick B. Hall}
\affil{Department of Astronomy, University of Toronto, 60 St. George Street,
Toronto, ON M5S~3H8, Canada}
\author{Marcin Sawicki}
\affil{Caltech, Mail Stop 320-47, Pasadena, CA 91125}
\author{C. J. Pritchet, F. D. A. Hartwick}
\affil{Department of Physics and Astronomy, University of Victoria, Victoria,
BC V8W~3P6, Canada}
\author{Aaron S. Evans}
\affil{Dept. of Physics and Astronomy, State University of New York at 
Stony Brook, Stony Brook, NY 11794-3800}

\begin{abstract}
We present new data on several $z=1.54$ radio-loud quasar fields 
from a sample of 31 at $z=1-2$ in which we have previously identified 
an excess population of predominantly red galaxies.
Narrow-band H$\alpha$ observations detect five candidate
galaxies at the quasar redshifts in three fields totaling 10.156\sq\arcmin,
a surface density $\sim$30 times higher than in previous surveys, 
even targeted ones.
SCUBA observations of three fields detect
at least one candidate quasar-associated galaxy.
Many galaxies with SEDs indicating considerable dust are not detected,
but the limits are only sufficient to rule out hyperluminous infrared galaxies.
Finally, quantitative photometric redshifts and SED fits are presented for one
``$J$-band dropout" galaxy with $J$$-$$K$$>$2.5 which is confirmed to be very
dusty ($E(B$$-$$V)$$\simeq$0.5) and background to the quasar at $\geq$99.9\%
confidence.
\end{abstract}


\section{Introduction}

It is of considerable interest to identify structures of galaxies
at z$>$1 to study the evolution of both galaxies and galaxy clusters.
Radio-loud quasars (RLQs) are obvious signposts around which to search for
clusters at z$>$1.  In Hall \& Green (1998; hereafter HG98) we presented
imaging of 31 RLQs at z=1--2 which revealed an excess population
of predominantly red galaxies.
Most candidate excess
galaxies' SEDs are consistent with them being at the quasar redshifts
and red due to high age or metallicity, but some are consistent with
being heavily dust-reddened galaxies, 
and/or background galaxies at z$>$2.5.
Here and in Hall {\it et\,al.} (1999) we present new observations of
these fields and further analyses of existing data which strengthen many of
our previous conclusions.

\section{Narrow-band H$\alpha$ Imaging}

In the past few years, narrow-band surveys have typically yielded a few
detections per survey of H$\alpha$ emitters at $z>1$ 
(see Teplitz, McLean \& Malkan 1999).
Our RLQ fields make promising targets for narrow-band 
searches for H$\alpha$ emission at the quasar redshifts.
We would hope to detect galaxies whose SEDs suggest they are dust-reddened 
and thus possibly actively star-forming and not to detect galaxies whose SEDs
suggest they are old and dust-free.

We observed Q~0835+580 with IRTF using a circularly variable filter (CVF),
Q~2149+212 with CFHT using a special narrow filter, and Q~2345+061 with both.
There is a $>3\sigma$ detection in each of the three fields
and two $3\sigma$ detections in the Q~2345+061 field.
Q~0835+580~(H$\alpha$1) is an unremarkable faint blue galaxy
with SFR$_{{\rm H}\alpha}$=14.7$\pm$2.5 $M_{\odot}$~yr$^{-1}$.
Based on $U$-band data for this field, we estimate
SFR$_{FUV}$=5.3$\pm$0.3 $M_{\odot}$~yr$^{-1}$,
in good agreement with SFR$_{{\rm H}\alpha}$ given the various uncertainties
(e.g. no correction for dust extinction has been made to either value).
None of the nine very red galaxies within 20\arcsec\ of Q~0835+580 were
detected in H$\alpha$,
strengthening the case for them being red due to age or
metallicity instead of dust (if they are at the quasar redshift).

In the Q~2345+061 field, the candidate H$\alpha$ emitter seen with IRTF 
(SFR$_{{\rm H}\alpha}$ = 4.9$\pm$0.1 $M_{\odot}$~yr$^{-1}$) is not confirmed
with CFHT, but the IRTF CVF is three times wider than the CFHT
narrow-band filter.  If the H$\alpha$ excess observed with IRTF is real,
the line must lie outside the CFHT filter.
Conversely, the two $3\sigma$ detections 
(SFR$_{{\rm H}\alpha}$=1.8$\pm$0.4 $M_{\odot}$~yr$^{-1}$) are seen with CFHT
but not IRTF.  Hoever, given the relative widths of the filters,
lines of the strength seen in the CFHT data could
be present but lost in the noise in the IRTF data.

Five detections over 10.156\sq\arcmin\ in these three 
fields (all among our top ten cluster candidates)
gives a surface density of 0.5$_{-0.2}^{+0.3}$~arcmin$^{-2}$, $\sim$30 times 
higher than previous shallower surveys, even targeted ones.
The deep CFHT H$\alpha$ images show that there are only three
quasar-associated galaxies with star formation rates of 
$>2~M_{\odot}~{\rm yr}^{-1}$ within fields $\sim$1~Mpc wide
centered on Q~2149+212 and Q~2345+061.
This is a lower limit which neglects extinction
and the velocity dispersion of the clusters,
but it illustrates the potential of deep wide-field
narrow-band data in studying star formation rates in high redshift clusters.

\section{Sub-millimeter Mapping}

The presence of a number of galaxies with SEDs strongly indicative
of substantial dust reddening in our RLQ fields (see HG98) suggested that
they might be detectable sub-mm sources.  Thus we observed the fields of
Q~0835+580, Q~1126+101, and Q~2345+061 with SCUBA on the JCMT.  
The reduced jiggle maps were cross-correlated with the beam map
and correlation coefficients measured; a high value was required 
to accept any potential source as real.

Q~2345+061 was detected at $2.8\sigma$, and Q~1126+101 at $3.4\sigma$.
Only one other source, dubbed Q~1126+101 (SM1), is securely detected,
but the limits on our relatively short exposures can only rule out
luminosities $\geq$10$^{13}$~L$_{\odot}$ for galaxies at the quasar redshifts.
Q~1126+101 (SM1) has two possible counterparts.
The closest 
is a candidate quasar-associated
red galaxy with $K=19.4$, $r-K=5.7$, $z-J=3.4$ and $J-K=2.7$.
The next closest 
has a moderate $r-K$ and
with $K=17.7$ is almost certainly foreground to the quasar.
Photometric redshifts and spectral types for the objects
may help determine which ID is most plausible.

\section{Photometric Redshifts and Spectral Types}

We are calculating photometric redshifts and spectral types
for objects in z$>$1 quasar fields with multicolor imaging data
to verify the existence of excess galaxies at the quasar redshifts or beyond,
investigate such galaxies' SEDs, and remove secure
foreground objects from consideration for future spectroscopy.
To compute photometric redshifts, 
a solar metallicity GISSEL model with synthetic Kurucz spectra
(Bruzual \& Charlot 1996) was calculated for ages 0--20~Gyr and
z=0--4 assuming either an instantaneous burst or a constant SFR and with dust
added using the Calzetti (1997) extinction law for ten values of $E(B$$-$$V)$
from 0 to 1.6.  Fluxes were computed and compared to observations 
to construct $\chi^2$ contour plots in age-$z$ space for each value of
$E(B$$-$$V)$ for each SFR scenario.

We present preliminary results on one object.  Q~1126+101 (425) was studied
since it is the brightest ``J-dropout'' ($J$$-$$K$$>$2.5) in its field and it
has a very red $z$$-$$J$$>$2.7.  For either a constant SFR or instantaneous
burst (the latter being shown in Figure 1), the lowest $\chi^2$ is given by
$E(B$$-$$V)$$\sim$0.5.  With $z_{ph}$=3.5$\pm$0.5, the object is background
to the $z$=1.54 quasar at $>$99.9\% confidence for all $E(B$$-$$V)$$<$0.7.  
For a constant SFR,
the very red colors of Q~1126+101 (425) require $E(B$$-$$V)$$>$0.5
at any $z$ to remain younger than the universe in any reasonable cosmology.
The required dust is consistent with its $J$$-$$K$ color being redder than
that of HR10, the prototypical dusty ERO (Extremely Red Object).
However, the prediction of HG98 that this particular object would be at the 
quasar redshift seems to be erroneous.  
Instead, it appears to be a member of the other
$J$-band dropout population proposed in HG98, namely galaxies at z$\geq$2.5
which have red $J$$-$$K$ colors due to the redshifted 4000\AA\ break.

\begin{figure}
\plotone{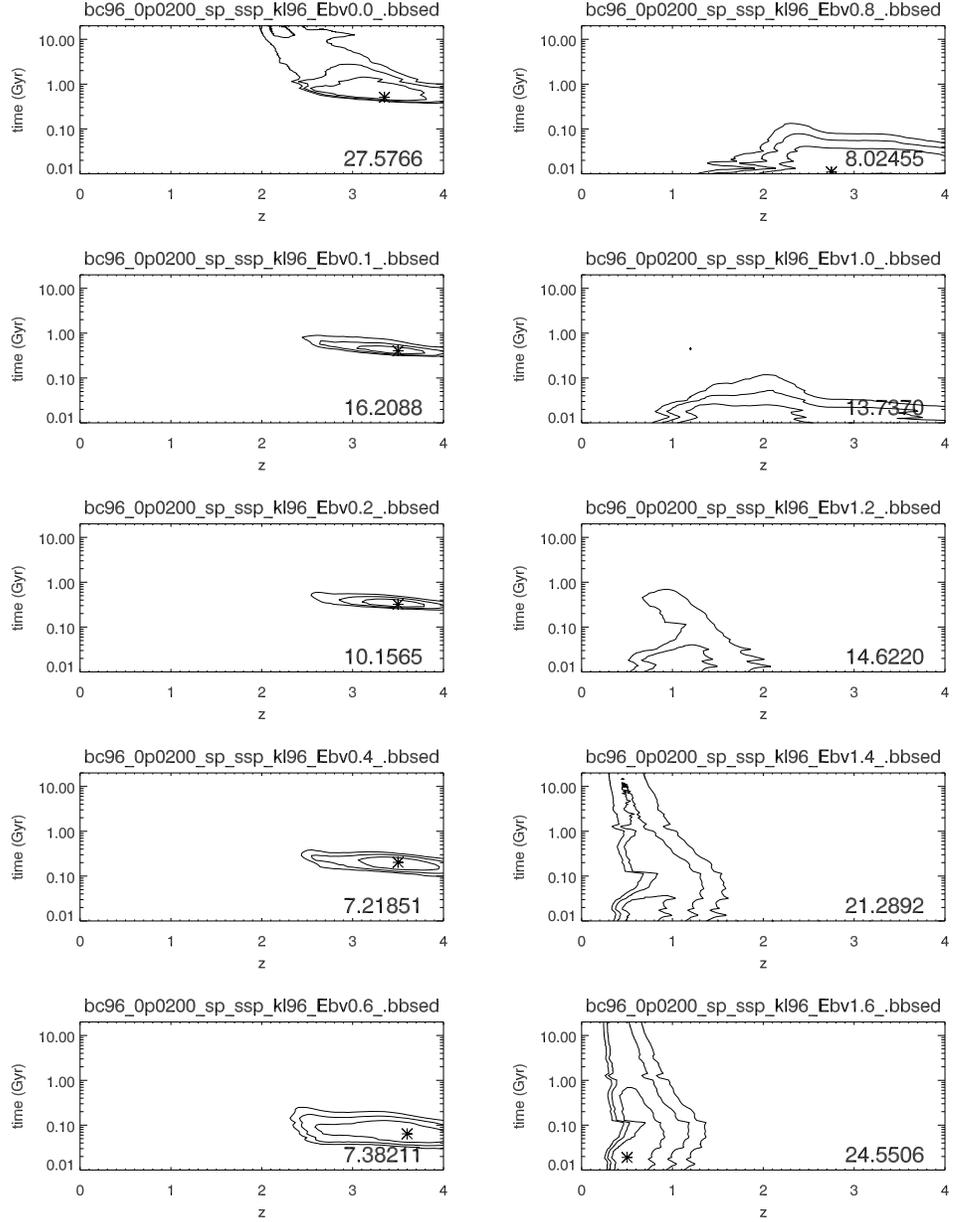}
\caption[Photometric Spectroscopy for Q~1126+101 (425)]{
$\chi^2$ contour plots for instantaneous burst fits
to the SED of object Q~1126+101 (425).
Each panel corresponds to a different $E(B-V)$, starting with 0 at top left and
increasing downward to 0.6 at the bottom of the first column and 1.6 at the
bottom of the second.
The asterisk shows the point with minimum $\chi^2$ in each panel.
From smallest to largest, the error contours enclose
the 90, 99 and 99.9\% probability regions respectively.
Note that the contours are drawn for each panel independently relative to the
minimum $unreduced$ $\chi^2$ for that panel,
which is printed in the bottom right corner.
}\label{fig_photz425.1126}
\end{figure}

\section{Conclusion}

These and other observations (Hall {\it et\,al.} 1999) 
are good evidence for galaxy overdensities around z$>$1 RLQs,
and for a population of red galaxies at z$\geq$2.5.
If confirmed by spectroscopy with 8-m class telescopes, it may prove
worthwhile to extend RLQ host cluster searches to z$>$2
by searching for ``$J$-dropout" galaxies.

\end{document}